\documentclass[11pt]{article}

\usepackage{amsmath,amssymb,latexsym,slashed}
\usepackage[numbers,sort&compress]{natbib}

\newcommand{\f}[2]{\frac{#1}{#2}}
\newcommand{\tf}[2]{{\textstyle \frac{#1}{#2}}}
\newcommand{\slaD}{\slashed{D}}
\newcommand{\de}{\partial}
\newcommand{\la}{\langle}
\newcommand{\ra}{\rangle}
\newcommand{\lla}{\la\!\la}
\newcommand{\rra}{\ra\!\ra}
\newcommand{\Oc}{O}
\newcommand{\Ac}{A}
\newcommand{\Pc}{P}
\newcommand{\Sc}{S}
\newcommand{\Gc}{{\cal G}}
\newcommand{\Rc}{{\cal R}}
\newcommand{\GG}{{\rm G}}
\newcommand{\RR}{{\rm R}}
\newcommand{\tr}{{\rm tr}\,}

\author{Matteo Giordano\footnote{giordano@bodri.elte.hu}
  \\ELTE E\"otv\"os Lor\'and University, Institute for
  Theoretical Physics,\\ P\'azm\'any P\'eter s\'et\'any 1/A, H-1117, Budapest,
  Hungary}
\title{Localised Dirac eigenmodes, chiral
  symmetry breaking, and Goldstone's theorem at finite temperature}

\date{}

\begin{document}

\maketitle

\begin{abstract}
  I show that a finite density of near-zero localised Dirac modes in
  the chirally broken phase of a gauge theory can lead to the
  disappearance of the massless excitations predicted by the Goldstone
  theorem at finite temperature.
\end{abstract}

\paragraph{Introduction}

Localisation has been the subject of intense research in condensed
matter physics since Anderson's seminal paper~\cite{Anderson:1958vr},
both on the theoretical and the experimental side (see
Ref.~\cite{anderson50} for a review).  The basic idea is that in the
presence of disorder the energy eigenstates of a quantum system can
become localised in space, i.e., mostly concentrated in a finite
spatial region whose size does not change as the system size
increases. In contrast, a delocalised mode extends throughout the
whole system, and keeps spreading out as the system size increases.
Anderson's original purpose was to show the absence of diffusion in
sufficiently disordered lattice systems, which leads to explain the
loss of zero-temperature conductance in a crystal with impurities in
terms of a disorder-induced metal-insulator transition (see the
reviews~\cite{thouless1974electrons,lee1985disordered,Evers:2008zz}).
In recent years the main focus has been on many-body localisation and
the resulting lack of thermalisation in closed quantum systems (see
the review~\cite{manybody}).

Surprisingly, localisation has been found also in a very different
setting, namely in the eigenmodes of the Dirac operator in gauge
theories at finite temperature~\cite{GarciaGarcia:2006gr,
  Kovacs:2009zj,Kovacs:2010wx,Kovacs:2012zq,Giordano:2013taa,Cossu:2016scb,
  Giordano:2016nuu,Kovacs:2017uiz,Holicki:2018sms,Giordano:2019pvc,
  Vig:2020pgq,Bonati:2020lal} (see Ref.~\cite{Giordano:2021qav} for a
recent review).  Gauge theories lie at the core of the Standard Model,
and are characterised by a local (gauge) invariance under some
symmetry group which largely dictates the form of the interactions,
mediated by the so-called gauge bosons.  In particular, strong
interactions are described by Quantum Chromodynamics (QCD), a gauge
theory with group SU(3) whose gauge bosons (``gluons'') mediate
interactions among spin-$\f{1}{2}$ fundamental fermions (``quarks'').
The phase diagram of QCD has been studied from first principles by
means of the nonperturbative lattice approach~\cite{Montvay:1994cy}.
At zero chemical potential and finite temperature $T$, QCD displays a
rapid (but analytic) crossover around
$T_c\approx 155\,{\rm MeV}$~\cite{Borsanyi:2010bp,Bazavov:2016uvm}
between a low-$T$, confined phase with quarks and gluons bound within
hadrons, and a high-$T$, deconfined phase where they are liberated in
the quark-gluon plasma (QGP).  Confining properties are determined by
the status of the approximate centre symmetry $\mathbb{Z}_3$: while
only explicitly (and mildly) broken at low $T$ by the presence of
quarks, it is also broken spontaneously (and strongly) at high $T$ by
the ordering of the Polyakov loop, i.e., the holonomy of the gauge
field along a straight path winding around the temporal direction.
The existence of the QGP has been confirmed experimentally, and the
study of its properties is the subject of extensive experimental
programs (see Ref.~\cite{Shuryak:2014zxa}).

The coupling of fermions and gauge bosons is encoded in the Dirac
operator in a gauge-field background, with physical observables
obtained integrating over gauge-field configurations. This is
analogous to ensemble averaging in a disordered system, with the Dirac
operator playing the role of the Hamiltonian, and the gauge-field
fluctuations that of a random interaction. One can then study the
spectrum and the localisation properties of the eigenmodes as with
disordered systems in condensed matter.  Most interesting are the
low-lying modes, which play an important role for the chiral
properties of the theory.  In the confined phase of QCD, a finite
density of near-zero modes signals the spontaneous
breaking~\cite{Banks:1979yr} of an approximate chiral symmetry.  At
high temperature in the deconfined phase this density vanishes and
chiral symmetry is effectively restored.  While in the low-$T$ phase
the low modes are delocalised, in the high-$T$ phase they are
localised up to a $T$-dependent ``mobility edge'', $\lambda_c$, in the
spectrum~\cite{Kovacs:2012zq,Cossu:2016scb,Holicki:2018sms}, where
they undergo a delocalisation (Anderson)
transition~\cite{Giordano:2013taa}.  The main source of disorder was
identified~\cite{Bruckmann:2011cc,Giordano:2015vla,Giordano:2016cjs}
with the fluctuations of the Polyakov loop around its ordered value;
this is supported by the critical properties of the Anderson
transition from localised to delocalised modes at
$\lambda_c$~\cite{Giordano:2013taa,Nishigaki:2013uya,Ujfalusi:2015nha}.

The connection between confinement, chiral symmetry breaking and
localisation is not fully understood yet.  In QCD neither centre nor
chiral symmetry is exact, and the transition is a crossover, so it is
not possible to make sharp statements.  For fundamental quarks, exact
centre symmetry is found in the ``quenched'' limit of infinite mass,
i.e., in pure gauge theory; exact chiral symmetry is found in the
opposite, ``chiral'' limit of massless quarks.  Investigations in a
clear-cut setting thus typically require separate studies of the
relation of localisation with deconfinement and with chiral
restoration.  Hints at a close relation between deconfinement and
localisation of the low Dirac modes come from SU(3) pure gauge theory,
which displays a first-order deconfining phase
transition~\cite{Boyd:1996bx}. In this theory localisation appears
precisely at the transition~\cite{Kovacs:2017uiz,Vig:2020pgq}.  The
same behaviour has been found also in other models with a sharp
deconfinement transition, both pure
gauge~\cite{Giordano:2019pvc,Bonati:2020lal,Baranka:2021san} and with
dynamical fermions~\cite{Giordano:2016nuu}.

Interestingly, in pure gauge SU(3) right above the deconfinement
temperature the density of near-zero modes is not
vanishing~\cite{Alexandru:2015fxa,Kovacs:2017uiz,Vig:2020pgq},
signalling in a loose sense the spontaneous breaking of chiral
symmetry by localised modes.  A similar peak of near-zero modes was
found also in QCD with near- and below-physical quark masses, near and
above the crossover temperature $T_c$, studying the overlap spectrum
on HISQ lattices~\cite{Dick:2015twa}, and their localised nature was
demonstrated for near-physical masses at $1.2T_c$ and $1.5T_c$.  This
peak may survive in the chiral limit~\cite{Kaczmarek:2021ser}, leading
to an intermediate regime with chiral symmetry broken by a condensate
originating from localised modes.

For massless adjoint fermions both centre and chiral symmetry are
exact, and the connection of localisation with deconfinement and
chiral symmetry restoration can be studied at once.  Numerical lattice
studies of the two-flavour case showed the presence of two distinct
phase transitions~\cite{Karsch:1998qj,Engels:2005te}: a deconfining
first-order one at $T_{\rm dec}$ (with a jump in the chiral
condensate), and a chirally-restoring second-order one at
$T_\chi>T_{\rm dec}$. In the intermediate range of temperatures
$T_{\rm dec}<T<T_\chi$ the nonzero chiral condensate implies a nonzero
density of near-zero modes. As centre symmetry breaking indicates that
the source of disorder is active, these modes are expected to be
localised; no direct study is, however, available.

Motivated by these findings, in this letter I discuss the possible
consequences of a nonzero density of near-zero localised modes in the
chiral limit at finite temperature.  The most dramatic scenario is the
disappearance due to localisation of the massless (Goldstone)
excitations associated with spontaneous symmetry breaking.  The idea
is not new, and has been put forward long ago in
Ref.~\cite{McKane:1980fs}, in the context of Anderson models, and in
Ref.~\cite{Golterman:2003qe}, in the context of the Aoki
phase~\cite{Aoki:1983qi} of quenched SU(3) gauge theory at $T=0$. The
present results are, however, new, as they concern relativistic
theories at finite temperature in a potentially physically relevant
setting.  I work in the imaginary-time, Euclidean path-integral
formulation, and in the continuum to avoid inessential
technicalities. The issues of regularisation and renormalisation of
ultraviolet (UV) divergences are discussed only briefly, as they do
not play any significant role.  A more detailed account will appear
elsewhere~\cite{giordano_GTshort}.

\paragraph{Finite-temperature gauge theories}

Consider a gauge theory with compact gauge group $G$ and $N_f$
degenerate flavours of fermions of mass $m$, transforming in some
representation of $G$ and minimally coupled to the gauge fields, at
finite temperature $T$. Euclidean time-ordered correlation functions,
denoted by $\la\ldots\ra$, are obtained as path integrals starting
from the partition function $Z$,
\begin{equation}
  \label{eq:partf}
    Z = \int [DB] e^{-S_{\rm g}[B]}\int [D\Psi D\bar\Psi]
    e^{-\int_\beta d^4x\, \bar\Psi (\slaD[B]  +   m)   \Psi }\,.
\end{equation}
Here integration is over gauge fields $B_\mu$ and Dirac fermi\-on
fields $\Psi$ and $\bar{\Psi}$, satisfying respectively periodic and
antiperiodic boundary conditions in the temporal direction, which is
compactified to a circle of extension $\beta=1/T$ (as indicated by the
subscript $\beta$). The gauge action $S_{\rm g}$ includes the usual
Yang--Mills and gauge-fixing terms, while the appropriate ghost terms
needed to restore gauge invariance~\cite{Bernard:1974bq} are included
in the integration measure. The gauge choice is ultimately irrelevant,
but a covariant gauge like Lorenz gauge makes all the relevant
spacetime symmetries manifest.  The (single-flavour) Dirac operator
reads $\slaD[B] \equiv \gamma_\mu (\de_\mu + igB_\mu)$, with
$\gamma_\mu$ the Euclidean, Hermitean Dirac matrices and $g$ the gauge
coupling.  Summation over repeated indices is understood.
Anti-Hermiticity of $\slaD$ and the chiral property
$\{\gamma_5,\slaD\}=0$ imply a purely imaginary spectrum, symmetric
about zero.  Thermal averages $\lla\ldots\rra_\beta$ of real-time
observables are re\-con\-struc\-ted from the Euclidean correlation
functions by Wick rotation back to Minkowski
spacetime~\cite{Bros:1996mw,Strocchi:2008gsa,Meyer:2011gj,Cuniberti:2001hm}.

For $m=0$ the fermionic action has a chiral symmetry
${\rm SU}(N_f)_L\times {\rm SU}(N_f)_R \sim {\rm SU}(N_f)_V\times {\rm
  SU}(N_f)_A$, explicitly broken down to its vector part
${\rm SU}(N_f)_V$ by a mass term. The starting point of the present
analysis is the following Ward--Takahashi (WT) identity associated
with ${\rm SU}(N_f)_A$ (see Ref.~\cite{giordano_GTshort} for a
detailed derivation),
\begin{equation}
  \label{eq:WT1}
  - \de_\mu\la \Ac^a_\mu(x) \Pc^b(0)\ra + 2m 
  \la \Pc^a(x)\Pc^b(0)\ra =
  \delta^{(4)}(x)\delta^{ab}\Sigma\,,
\end{equation}
where $\Sigma \equiv \tf{1}{N_f}\la \Sc(0)\ra$ is the chiral
condensate, and the fact that ${\rm SU}(N_f)_V$ is not spontaneously
broken~\cite{Vafa:1983tf} has been used.  Here
$\Ac^a_\mu\equiv\bar{\Psi}\gamma_\mu\gamma_5 t^a\Psi$ is the
flavour-nonsinglet axial current,
$\Pc^a\equiv\bar{\Psi}\gamma_5 t^a\Psi$ the nonsinglet pseudoscalar
density, and $\Sc\equiv\bar{\Psi} \Psi$ the singlet scalar density,
with $t^a$ the Hermitean generators of ${\rm SU}(N_f)$, normalised to
$\tr t^a t^b= \f{1}{2}\delta^{ab}$.  Eq.~\eqref{eq:WT1} is one of the
infinite set of WT identities expressing the partial conservation of
the axial current (see, e.g., Ref.~\cite{Brandt:2014qqa}), and the
starting point for the derivation of the Gell-Mann--Oakes--Renner
relation between the light-quark and pion masses in the low-$T$ phase
of QCD (see Ref.~\cite{Brandt:2014qqa}).

WT identities may in general be spoiled by the renormalisation
procedure.  Here, using the lattice regularisation and
Ginsparg--Wilson fermions~\cite{Ginsparg:1981bj}, one can show
nonperturbatively that this is not the case, whenever a continuum
limit can be defined.  An exact lattice chiral
symmetry~\cite{Luscher:1998pqa} guarantees~\cite{Hasenfratz:1998jp}
that $m$ renormalises only multiplicatively; that $\Ac^a_\mu$ requires
no further renormalisation after the usual mass, coupling and field
renormalisation; and that the multiplicative renormalisation constants
$Z_P$ and $Z_S$ of the pseudoscalar and scalar densities satisfy
$Z_P=Z_S=Z_m^{-1}$ with $Z_m$ the mass renormalisation constant. It
also implies that all additive divergent contact terms drop from
Eq.~\eqref{eq:WT1} in the chiral limit. The WT identity
Eq.~\eqref{eq:WT1} can then be treated as a meaningful relation
between finite, renormalised, continuum quantities.

\paragraph{Goldstone theorem at finite temperature}

The WT identity Eq.~\eqref{eq:WT1} can be used to provide a (as far as
I know, novel) derivation of Goldstone's theorem at finite
temperature~\cite{Lange:1965zz,Kastler:1966wdu,Swieca:1966wna,
  Morchio:1987wd,Strocchi:2008gsa}.  In (Euclidean) energy-momentum
space one finds
\begin{equation}
    \label{eq:WT5}
  i\omega_n \Gc_{4}(\omega_n,\vec{p}\,) +
    i p_k\Gc_k(\omega_n,\vec{p}\,)+
  \Rc(\omega_n,\vec{p}\,)= \Sigma\,, 
\end{equation}
where using vector-flavour invariance I have set
\begin{align}
  \label{eq:ft_corr}
  \delta^{ab}   \Gc_{\mu}(\omega_n,\vec{p}\,) &\equiv \int_\beta d^4x\,
    e^{i(\omega_n t  + \vec{p}\cdot\vec{x})}\la \Ac^a_\mu(x)\Pc^b(0)\ra\,,\\
  \label{eq:ft_corr2}
  \Rc(\omega_n,\vec{p}\,) &\equiv \int_\beta
    d^4x\, e^{i(\omega_n t + \vec{p}\cdot\vec{x})}\, R(x)\,,\\
 \delta^{ab} R(x) &\equiv  2m\la \Pc^a(x) \Pc^b(0) \ra   \,.
\end{align}
Periodicity restricts Euclidean energies to the discrete Matsubara
frequencies $\omega_n = \f{2\pi n}{\beta}$, $n\in \mathbb{Z}$. Using
the relation between Euclidean correlators and real-time thermal
expectation values~\cite{Bros:1996mw,Meyer:2011gj,Cuniberti:2001hm}
together with relativistic locality one shows that
$\lim_{\vec{p}\to 0}p_k \Gc_k(\omega_n,\vec{p}\,)=0$ for $n\neq 0$.
Invariance under the time reflection $t\to\beta-t$ implies
$\Gc_4(-\omega_n,\vec{p}\,) = -\Gc_4(\omega_n,\vec{p}\,)$, and so
$ \Gc_4(0,\vec{p}\,)=0$.  The same symmetry implies
$\Rc(-\omega_n,\vec{p}\,)=\Rc(\omega_n,\vec{p}\,)$.

Setting now $\GG(\omega_n)\equiv\Gc_4(\omega_n,\vec{0}\,)$, denoting
quantities in the chiral limit by the subscript ``$*$'', and under the
usual assumption that $R\to 0$, from Eq.~\eqref{eq:WT5} one finds in
the chiral limit
\begin{equation}
  \label{eq:WT6}
  i\omega_n \GG_*(\omega_n) = \Sigma_*\,, \quad n\neq 0\,.
\end{equation}
This implies that $\GG_*$ can be continued analytically (in the sense
of the unique Carlsonian
interpolation~\cite{Bros:1996mw,Cuniberti:2001hm}) to the function
$\bar\GG_*(\Omega)=\Sigma_*/(i\Omega)$ of the complex variable
$\Omega$, from which one recovers the corresponding real-time thermal
correlation function.  Setting (for $\omega\in\mathbb{R}$)
\begin{equation}
  \label{eq:APcomm}
  \delta^{ab}
  c(\omega) \equiv  \lim_{\vec{p}\to 0} 
  \int d^4 x\,
  e^{i(\omega t - \vec{p}\cdot \vec{x})}
  \lla
  \,[\hat{\Ac}^a_0(x),\hat{\Pc}^b(0)] \,\rra_\beta\,,
\end{equation}
where $\hat{\Ac}^a_\mu$ and $\hat{\Pc}^a$ are the Minkowskian
axial-vector current and pseudoscalar density {\it operators}, one
finds~\cite{Bros:1996mw,Meyer:2011gj,Cuniberti:2001hm}
\begin{equation}
  \label{eq:APcomm2}
  i  c(\omega) = 
  \bar\GG(0^+-i\omega)
  -\bar\GG(0^--i\omega)\, .
\end{equation}
From Eq.~\eqref{eq:WT6} one then has in the chiral limit
\begin{equation}
  \label{eq:APcomm3_0}
 i c_*(\omega) =  \f{\Sigma_*}{\omega+i0^+}-\f{\Sigma_*}{\omega+i0^-}
= -2\pi i\Sigma_*\delta(\omega)\,.
\end{equation}
When $\Sigma_*\neq 0$, this implies the presence of quasi-particle
excitations with zero energy at zero momentum, i.e., the Goldstone
theorem at finite temperature~\cite{Strocchi:2008gsa}.

More generally, when $R\neq 0$, after analytic interpolation (in the
Carlsonian sense) of $\RR(\omega_n)\equiv \Rc(\omega_n,\vec{0})$ to
the function $\bar\RR(\Omega)$, one finds
$\bar\GG(\Omega)=(\Sigma-\bar\RR(\Omega))/(i\Omega)$, and
Eq.~\eqref{eq:APcomm2} gives
\begin{equation}
  \label{eq:APcomm3}
    i c(\omega)  = -2\pi i \left[\Sigma - \bar\RR(0^+)\right] \delta(\omega)
    -  \f{\bar\RR(0^+-i\omega)-\bar\RR(0^++i\omega)}{\omega}
    \,,
\end{equation}
where the relation $\bar\RR(0^--i\omega) =\bar\RR(0^++i\omega)$
following from the symmetries of $\Rc$ has been used.  The second term
in Eq.~\eqref{eq:APcomm3} is regular at $\omega=0$, while the
$\delta$-term, and so Goldstone quasi-particles, are present if
$\Sigma - \bar\RR(0^+)\neq 0$.

\paragraph{Pseudoscalar correlator in the chiral limit}

The standard assumption $R\to 0$ is based on assuming that
$\la \Pc^a(x) \Pc^b(0) \ra$ is regular enough as a function of $m$ in
the chiral limit. I show now that this in fact may not be the case
when a finite density of localised near-zero modes is present:
$\la \Pc^a(x) \Pc^b(0) \ra$ can develop a $1/m$ infrared divergence
that compensates the factor of $m$, leading to a finite remnant $R_*$.
This mechanism is similar to that discussed in
Ref.~\cite{Golterman:2003qe}. Notice that the limiting procedure used
here cannot be avoided to study symmetry breaking in the chiral limit
(unless one uses chirally-violating boundary conditions), as setting
$m=0$ in a finite volume automatically enforces chiral symmetry.

In a finite spatial volume $V$, where the eigenvalues $i\lambda_n$ of
$\slaD$ are discrete, the {\it unrenormalised} pseudoscalar correlator
reads
\begin{equation}
  \label{eq:psccorr1}
  \la \Pc_B^a(x)\Pc_B^b(0)\ra
  =-\f{\delta^{ab}}{2} \left\la\sum_{n,n'}
    \f{\Oc^{\gamma_5}_{n'n}(x)\Oc^{\gamma_5}_{nn'}(0)}{(i\lambda_n
    +m_B)(i\lambda_{n'} +m_B)}\right\ra  =
    -\delta^{ab} \Pi_B(x)\,,
\end{equation}
where $\Pc_B^a$ and $m_B$ are the {\it bare} pseudoscalar density and
mass, and the sums are restricted to $|\lambda_{n,n'}|\le \Lambda$ for
UV regularisation purposes.  Here
$\Oc^\Gamma_{nn'}(x) \equiv (\psi_n(x),\Gamma\psi_{n'}(x))$, where
$\slaD\psi_n=i\lambda_n\psi_n$ with $\psi_n$ normalised to 1 and the
scalar product $(\cdot,\cdot)$ involves only Dirac and gauge-group
indices. All contributions to Eq.~\eqref{eq:psccorr1} irrelevant in
the limit $V\to\infty$, followed by the chiral limit $m\to 0$, will
now be dropped.  The number of exact zero modes grows only like
$\sqrt{V}$, and so they can be neglected as $V\to\infty$.  In the
chiral limit, modes outside of an infinitesimal neighbourhood of the
origin will give at most a finite contribution.  One can then replace
$\Lambda$ with an arbitrary cut-off, as long as it is kept
non-vanishing in the chiral limit. In particular, this removes
possible sources of additive UV divergences, $\Pi^{\rm add.\, div.}$.
Including the renormalisation factor $Z_P^{-2}=Z_m^2$, one then
replaces the term on the right-hand side of Eq.~\eqref{eq:psccorr1}
with the {\it renormalised} quantity
$\Pi(x) = Z_m^2[\Pi_B(x)-\Pi^{\rm add.\, div.}(x)]$,
\begin{equation}
  \label{eq:psccorr1_ter_ren}
  \Pi(x) =  \f{1}{2}
  \int_{-\mu}^{\mu} d\lambda \int_{-\mu}^{\mu} d\lambda' \,\f{P(\lambda,
    \lambda'; m;x)}{(i\lambda+m)(i\lambda'+m)} + \ldots\,,
\end{equation}
where\footnote{Finiteness of $P$ is proved in
  Ref.~\cite{giordano_GTshort} using the methods of
  Ref.~\cite{DelDebbio:2005qa}.}
\begin{equation}
  \label{eq:ppcorf_ren}       P(\lambda,\lambda';m;x)
  \equiv \left\la\sum_{\lambda_{n,n'}\neq
      0}
    \delta\left(\lambda-\f{\lambda_n}{Z_m}\right)
    \delta\left(\lambda'-\f{\lambda_{n'}}{Z_m}\right)
    \Oc^{\gamma_5}_{n' n}(x)\Oc^{\gamma_5}_{nn'}(0)\right\ra 
  \,,
\end{equation}
$m=Z_m^{-1}m_B$ is the {\it renormalised} mass, $\mu$ is a finite mass
scale, and dots stand for terms negligible in the chiral limit.  A
change of variables $\lambda^{(\prime)}=mz^{(\prime)}$ should convince
the reader that a $1/m$ divergence can only originate from the terms
with $\lambda_n=\pm \lambda_{n'}$ in Eq.~\eqref{eq:ppcorf_ren}, while
the rest can give at most a $(\log m)^2$ divergence.  Exploiting also
the symmetry of the spectrum one then replaces
Eq.~\eqref{eq:psccorr1_ter_ren} with\footnote{Accidental degeneracies
  of nonzero eigenvalues appear on a set of configurations of zero
  measure and can be neglected.}
\begin{equation}
  \label{eq:psccorr1_ter_ren2}
  \Pi(x)=  
  \int_{0}^{\mu} d\lambda \left(\f{C^1(\lambda;m;x)}{\lambda^2+m^2}
    + \f{(m^2-\lambda^2)C^{\gamma_5}(\lambda;m;x)}{(\lambda^2+m^2)^2}
  \right) + \ldots\,,
\end{equation}
where                               
\begin{equation}
  \label{eq:psccorr3_alt2}   C^\Gamma(\lambda;m;x)\equiv
\left\la \sum_{\lambda_{n}\neq      0}
\delta\left(\lambda-\f{\lambda_n}{Z_m}\right) 
  \Oc^{\Gamma}_{nn}(x)\Oc^{\Gamma}_{nn}(0)\right\ra\,.
\end{equation}
The fate of $C^\Gamma$ in the chiral limit is determined by the
localisation properties of the eigenmodes.  For eigenmodes spread out
on the whole space, the local density $\Oc^{1}_{nn}(x)$ is
approximately uniform, $\Oc^{1}_{nn}(x)\sim 1/(\beta V)$, and the same
is expected for $\Oc^\Gamma_{nn}(x)$ as well. Their contribution to
Eq.~\eqref{eq:psccorr3_alt2} is then expected to vanish in the
thermodynamic limit: one expects qualitatively
$\Oc^\Gamma_{nn}(x)\sim 1/(\beta V)$, but only $O(V)$ terms in the
sum, so that $C^\Gamma\to 0$ as $V\to \infty$.  On the other hand, a
localised mode is essentially concentrated in some region of finite
spatial size $V_0$, so that $\Oc^\Gamma_{nn}(x)\sim 1/(\beta V_0)$ for
$x$ inside that region, and practically zero outside. Due to
translation invariance, the probability (defined over the ensemble of
gauge configurations) that $x=0$ belongs to the localisation region is
$V_0/V$. If there is a finite density of localised modes around
$\lambda$, one finds a contribution of order
$V\cdot V_0/V \cdot 1/V_0^2=1/V_0$, and so a non-zero $C^\Gamma$ is
expected (and possible only) in spectral regions with localised
modes.\footnote{For sharply localised modes $C^\Gamma$ would vanish
  for $|\vec{x}|$ larger than the typical localisation length; as
  localisation is typically exponential, an exponential suppression in
  $|\vec{x}|$ is expected instead.}  The same argument shows that
modes with nontrivial fractal dimension, i.e., concentrated in regions
of size $V_0\propto V^\alpha$ with $0<\alpha<1$, do not contribute to
$C^\Gamma$ in the thermodynamic limit.

I now assume that localised modes are present in a spectral region
near $\lambda=0$, up to a mobility edge $\lambda_c(m)$.  This is the
situation observed in several gauge theories at sufficiently high
temperature for nonzero quark
masses~\cite{Kovacs:2010wx,Kovacs:2012zq,Cossu:2016scb,Kovacs:2017uiz,
  Vig:2020pgq,Giordano:2019pvc,Bonati:2020lal,Baranka:2021san}.  If
other spectral regions contain localised modes, I assume that they
remain separated from the origin in the chiral limit.  The
UV-finiteness of $P$ in Eq.~\eqref{eq:ppcorf_ren} implies that
$\lambda_c$ renormalises like a quark mass, and so $\lambda_c/m$ is
renormalisation-group invariant, as already suggested in
Ref.~\cite{Kovacs:2012zq}.  It is now straightforward to evaluate
Eq.~\eqref{eq:psccorr1_ter_ren2} in the chiral limit, and find for
(renormalised) $R_*(x)$
\begin{equation}
  \label{eq:remnant_x}
- R_*(x) = 
  \xi\pi C_{\rm loc}^1(0;0;x) + \eta C_{\rm loc}^{\gamma_5}(0;0;x)\,.
\end{equation}
Here $\xi \equiv\f{2}{\pi}\arctan \kappa$,
$\eta\equiv\f{2\kappa}{1+\kappa^2}$,
$\kappa\equiv\lim_{m\to 0}\f{\lambda_c}{m}$, and ``loc'' denotes the
restriction of the spectral sums to localised modes only. Here and
below, the following order of limits is understood,
$f(0;0) \equiv \lim_{m\to 0}\lim_{\lambda\to 0}\lim_{V\to \infty}
f(\lambda;m)$.

\paragraph{Localisation and Goldstone modes}

The localised nature of the modes contributing to $R_*(x)$ allows one
to exchange spacetime integration in the zero-momentum limit with the
chiral limit and the infinite-volume limit to get
\begin{equation}
  \label{eq:remnant_p}
  \RR_*(0) =   
  \int_\beta d^4x \, R_*(x) 
  = - \xi\pi \rho_{\rm loc}(0;0)\,,
\end{equation}
where $\rho_{\rm loc}$ is the restriction to localised modes of the
spectral density,
\begin{equation}
  \label{eq:specdens}
  \rho(\lambda;m) \equiv \lim_{V\to\infty}\f{1}{\beta V}\left\la
  \sum_{\lambda_{n}\neq      0}
  \delta\left(\lambda-\f{\lambda_n}{Z_m}\right) \right\ra\,.
\end{equation}
Using the Banks-Casher relation
$\Sigma_*=-\pi\rho(0;0)$~\cite{Banks:1979yr}, and assuming that
$\RR_*(0)=\bar\RR_*(0^+)$,\footnote{This would not be the case in the
  presence of a transport peak $\propto\omega\delta(\omega)$ in the
  spectral function~\cite{Meyer:2011gj}. However, such a term is not
  expected in the pseudoscalar
  channel~\cite{Karsch:2003wy,Aarts:2005hg,Burnier:2017bod}.}  one
obtains from Eq.~\eqref{eq:APcomm3}
\begin{equation}
  \label{eq:remnant_final}
  c_*(\omega)|_{\rm singular} = 2\pi^2 [\rho(0;0)-\xi\rho_{\rm
    loc}(0;0)]\delta(\omega)\,. 
\end{equation}
The fate of the Goldstone excitations depends on the combination
$\rho(0;0)-\xi\rho_{\rm loc}(0;0)$: this is the main result of this
letter. In typical disordered systems, localised and delocalised modes
do not coexist in the same spectral region, so if modes near the
origin are delocalised one has $\rho_{\rm loc}(0;0)=0$, and the
standard situation arises.  If near-zero modes are localised then
$\rho(0;0)=\rho_{\rm loc}(0;0)$, and no Goldstone modes are present if
$\rho_{\rm loc}(0;0)=0$. If, instead, near-zero modes are localised
and $\rho_{\rm loc}(0;0)\neq 0$, one has three possibilities depending
on the value of $\kappa$: if $\kappa=0$ ($\lambda_c$ vanishes faster
than $m$ in the chiral limit), one recovers the standard result; if
$0<\kappa<\infty$ ($\lambda_c$ vanishes as fast as $m$), one still
finds Goldstone quasi-particles, although the coefficient of the
singular term is reduced with respect to the standard case; if
$\kappa=\infty$ ($\lambda_c$ vanishes more slowly than $m$ or remains
finite), then the Goldstone quasi-particles disappear. One would then
have chiral symmetry breaking without Goldstone excitations.

A few comments are in order to avoid misunderstandings.  Goldstone's
theorem is of course not violated or disproved, but simply evaded. In
fact, a nonzero contribution $R_*(x)$ to the WT identity in the chiral
limit indicates that the relevant current is not conserved, and so the
main assumption of the theorem does not hold.  A nonzero density of
near-zero modes (if $\kappa>0$) leads effectively to an explicit
breaking of the symmetry in the chiral limit, in a way reminiscent of
the formation of anomalies (although here in the IR rather than UV
regime).  In the general case, this accounts only partially for the
full symmetry breaking as measured by the condensate, and spontaneous
breaking is also present. However, if $\kappa=\infty$ all the breaking
effect is explicit, and no massless excitation is present.

\paragraph{Acknowledgments}
I thank C.~Bonati, M.~D'Elia, S.~D.~Katz, T.~G.~Ko\-v\'acs,
D.~N\'ogr\'adi, A.~P\'asztor, A.~Portelli, and Zs.~Sz\'ep for useful
discussions.  This work was partially supported by the NKFIH grant
KKP-126769.

\paragraph{}

\bibliographystyle{h-physrev_mod}
\bibliography{references_gt}

\end{document}